\documentclass[sigchi-a]{acmart}
\usepackage{hyperref}
\usepackage{subfigure}
\AtBeginDocument{%
  \providecommand\BibTeX{{%
    \normalfont B\kern-0.5em{\scshape i\kern-0.25em b}\kern-0.8em\TeX}}}

\setcopyright{acmcopyright}
\copyrightyear{2018}
\acmYear{2018}
\acmDOI{10.1145/1122445.1122456}

\acmConference[GROUP 2020]{GROUP 2020: ACM Conference on Supporting Groupwork}{January 06--08, 2020}{Sanibel Island, FL}
\acmBooktitle{GROUP 2020: ACM Conference on Supporting Groupwork,
  January 06--08, 2020, Sanibel Island, FL}
\acmPrice{15.00}
\acmISBN{978-1-4503-9999-9/18/06}

\begin{document}
\title{Willing Buyer, Willing Seller: Personal Data Trade as a Service}

\author{Lindah Kotut}
\affiliation{%
  \institution{Department of Computer Science}
  \institution{Virginia Tech}
  \city{Blacksburg} 
  \state{Virginia} 
  \postcode{24060}
}
\email{lkotut@vt.edu}

\author{Timothy L. Stelter}
\affiliation{%
  \institution{Department of Computer Science}
  \institution{Virginia Tech}
  \city{Blacksburg} 
  \state{Virginia} 
  \postcode{24060}
}
\email{tstelter@vt.edu}

\author{Michael Horning}
\affiliation{%
  \institution{Department of Communications}
  \institution{Virginia Tech}
  \city{Blacksburg} 
  \state{Virginia} 
  \postcode{24060}
}
\email{mhorning@vt.edu}

\author{D. Scott McCrickard}
\affiliation{%
  \institution{Department of Computer Science}
  \institution{Virginia Tech}
  \city{Blacksburg} 
  \state{Virginia} 
  \postcode{24060}
}
\email{mccricks@cs.vt.edu}

\renewcommand{\shortauthors}{Kotut, Horning and McCrickard}

\begin{abstract}
There is an increased sensitivity by people about how companies collect information about them, and how this information is packaged, used and sold. This perceived lack of control is highlighted by the helplessness of users of various platforms in managing or halting what data is collected from/about them. In a future where users have wrested control of their data and have the autonomy to decide what information is collected, how it is used and most importantly, how much it is worth, a new market emerges. This design fiction considers possible steps prescient companies would take to meet these demands, such as providing third-party subscription platforms offering personal data trade as a service. These services would provide a means for transparent transactions that preserve an owner's control over their data; allowing them to individually make decisions about what data they avail for sale, and the amount of compensation they would accept in trade.  
\end{abstract}


\maketitle

\section{Data Subscription Service: An Introduction}
With the increased digitization of personal information, companies behind mobile applications and online services collect users' data for profit. In hiking communities for example, these practices are useful primarily because of the hikers' unique data needs and  also the wide variety of data that they generate while on the trail. By companies packaging and reselling these types of data, the hikers from whom the data was harvested have no understanding and control over what information is exactly collected, who it is sold to and how it is used past the selling. 
 
Hikers' data are useful for variety of applications: mapping terrain, reporting weather conditions, measuring body performance etc. Applications that assist the hiker in achieving these goals also tend to package the hiker's data for sale to other companies. The monetary or equivalent benefit the company receives as a result of collected data is not passed back to the hiker. 

\begin{marginfigure}
\includegraphics[width=0.7\linewidth]{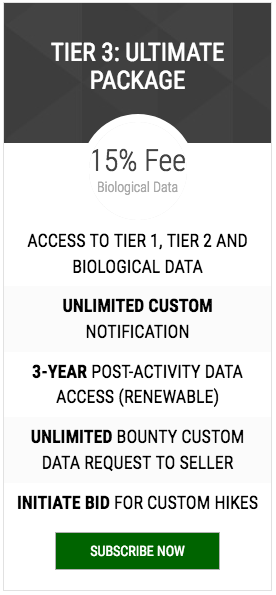}
\centering
\caption{Conceptual model of data packages presented as a tier. In this tier, a buyer can purchase access to environmental, situational and biometric data. A 15\% fee is applied by the company to offset the cost collecting, warehousing, processing and disseminating data and insights. Figure \ref{fig:full-tier} showcases how the rest of the tiers will be packaged}
\label{fig:access-tiers}
\end{marginfigure}

There is however an increased awareness of how the user has no control over personal data, knowing they are the commodity for sale \cite{Fiesler2018}. This increased knowledge and understanding of how users are a cog in the wheel of ``surveillance capitalism''\cite{Lin2018} has shown to lead users to abandon those services that collect the data about them in search of those services that allow them to retain personal control over their data \cite{Morris2019}. This trend will only continue as more users disagree with how their data  is collected and used. As this trend shows, there is a need for platforms that allow users to control information collected about them and provides absolute transparency of how their information is used. 

Our subscription service goes a step further, we allow the users to directly and explicitly \textit{sell} their data. As the ``free'' services that turn users into products have shown, there is a market for data. We remove the middle-man in the transaction and allow the user to \textit{sell} their information directly to buyers. Depending on perceived need and the amount of compensation, we allow the user (i.e. the seller) to determine whether it is worth a trade of personal information \cite{Liu2013}.

\section{Addressing Need (Product Design)}
'
Using a popular hiker saying \textit{``The Trail Provides''} \cite{TheTrailProvides} used to denote that a hiker's need will be met when they most need it, we propose a data sharing and incentive model. This model relies on \textit{``willing buyer, willing seller''} revised policy that allow buyers to bid for data they wish to know, and/or also allowing the seller to set the price for data they produce, such that both the buyer and the seller are able to exchange data at the fair market rate \cite{Slade2019}. Using \textit{The Trail Provides} principle, we can showcase the benefit that our subscription provides to both parties in the trade. A platform to sell data allows the sellers (hikers) to receive compensation for data and also ease the cost of current or future hikes, while at the same time satisfy pre-hikers' (buyers) need for nuanced and contextual information to inform their own curiosity about the trail, or to use to inform their own future hikes.

\subsection{The Incentive model}
Our subscription service offers the option to package the data for sale. For convenience, we provide initial data tiers that the sellers could adopt and allow the buyers to purchase subscriptions for access. We use a simple pricing formula:  the more intimate the data, the more expensive it is; the more difficult it is to obtain the data, the more expensive it is. These considerations allow for both a fair return on investments to the sellers and as an incentive model for buyers to access only what they need. Figure \ref{fig:access-tiers} is a typical example of how a tier is packaged, while Figure \ref{fig:full-tier} provides the entire data packaging that we offer. 

The seller, such as a hiker, also has the option to set her or his own price, or use recommended pricing depending on market sources. This can be done before the hike commences, during the hike, or even at the end of hike. We also provide a means for third parties to also supply \textit{milestone funds} to incentivize sellers to achieve a set milestone to complete, that is in line with the companies' needs (e.g. sensor data to measure shoe performance for a 500-mile hike). 

\begin{marginfigure}
\includegraphics[width=\linewidth]{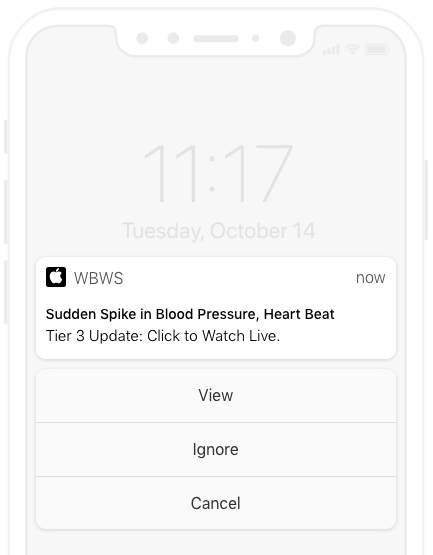}
\centering
\caption{Live Push notification customized to alert for unexpected events}
\label{fig:push-notification}
\end{marginfigure}

\subsection{Data Sources and Data Integrity}
In order to preserve data integrity of real-time data \cite{Sannon2018}, our service automatically collects from the seller and uploads to our cloud servers at frequent intervals making it easy for the seller to share the data with subscribers (we also provide sensors for rent for a small fee). Our system also provides automatic visualization so that the buyers can either view real-time information on a dashboard (Figure \ref{fig:full-dashboard}) or set alerts that will automatically notify them in real-time whenever a triggering event occurs (Figure \ref{fig:push-notification}).

Our company also has sensors and specialized equipment that the seller can purchase or rent. Our sensors use standard universal APIs to enable plug-and-play uses by sponsors and to account for atypical sensors. Figure \ref{fig:sensor-placement} shows different placement of sensors that would automatically collect and disseminate data without interrupting the hiker's pace and/or trail experience. 

While on the hike, light-weight, and non-power-intensive sensors supply data to a connected SIM-enabled command-and-control hub, that subsequently uploads this data at set intervals or streamed live. If connection is lost, uploads would proceed automatically once the connection is re-established. This makes it easier for the seller not to worry about missing updates, while allowing the data to receive just-in-time, reliable information. 

\subsection{Dashboard}
Our subscription service provides a seamless interface between the buyer and the seller. We are transparent in how we make money: from sellers, by fees from rented equipment; from buyers, in the form of subscription fees (Figure \ref{fig:access-tiers}) to process payment, store data and provide just-in-time updates.  Importantly, our service provides a means for the the sellers to receive continuous income for data that would otherwise be forgotten after long-term disuse \cite{Khan2018}. 

\begin{figure*}%
\centering
\subfigure[Dashboard allowing both the buyer and the seller to manage data]{%
\label{fig:full-dashboard}%
\includegraphics[angle=90,width=0.45\linewidth]{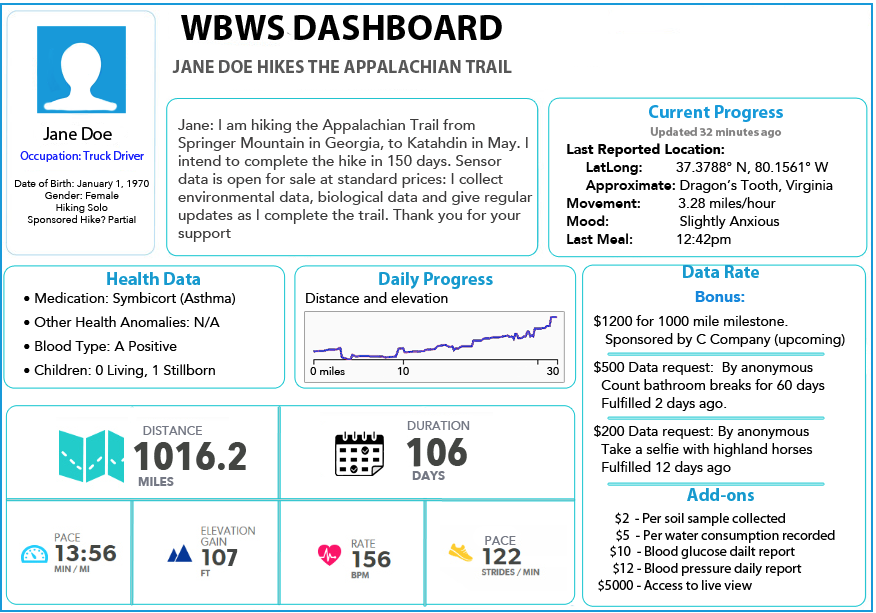}}%
\qquad
\subfigure[Data packaged to be sold in three tiers by order of sensitivity]{%
\label{fig:full-tier}%
\includegraphics[angle=90,width=0.45\linewidth]{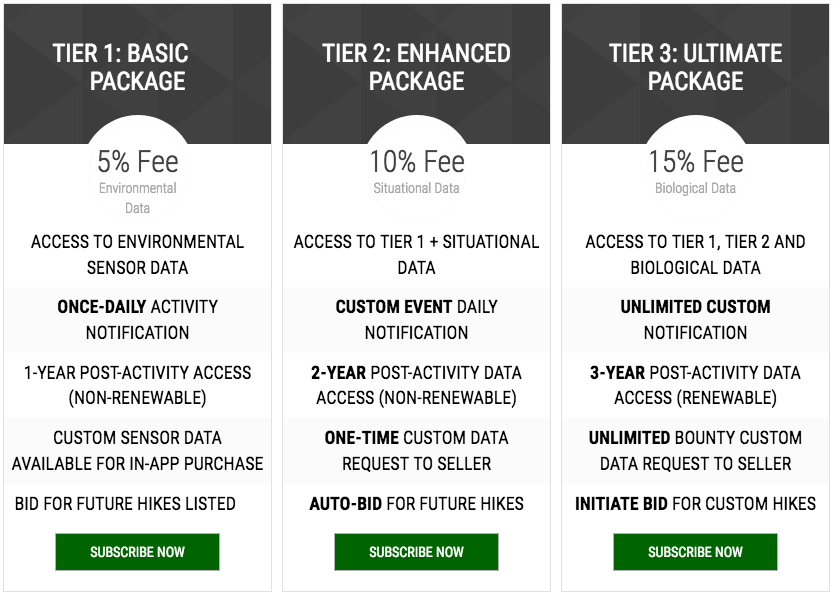}}%
\caption{Data pricing, data packaging and data management}
\label{fig:full-concept}
\end{figure*}

Both the sellers and the buyers have access to the dashboard where the bidding process for data and terms of use are set and agreed upon. The dashboard provides a visual of the escrow service we provide, to allow for a trustworthy stewardship that also provides for a seamless release of funds at milestone completion that is automatically updated and triggered from the sensor data we receive from the seller. Figure \ref{fig:full-dashboard} showcases a buyer's view of the data on a dashboard. 

\begin{marginfigure}
\includegraphics[width=\linewidth]{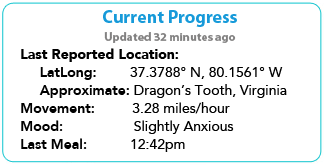}
\centering
\caption{A snippet of real-time sensor data viewed from the dashboard that follows the principle of \textit{Quantified Self}\cite{QuantifiedSelf}. The data includes exact and approximated location, the hiker's pace, the hiker's mood and updates about meals. Figure \ref{fig:full-dashboard} showcases the entire dashboard.}
\label{fig:dashboard-snippet}
\end{marginfigure}

Both the buyer and the seller can see all the personal information about the seller according to the tier the buyer has subscribed to, such as real-time location updates, last meal eaten, the number of bathroom breaks taken, the elevation gains, the pace etc. The dashboard also provides convenience by offering visualization options for data received from the seller. Figure \ref{fig:dashboard-snippet} provides a dashboard snippet that shows how both distance and time can be viewed to provide context on the progress of a seller who is in the process of completing their hike.

To be in the know, the buyer can simply set notification for events of interest that can be received as a notification (Figure \ref{fig:push-notification})
 with prompts to monitor events live as they happen. We provide add-on services that  follow the concept of in-app purchases prevalent in mobile apps. This provides the seller with an additional income stream by setting access to some additional information for a premium, which allow the buyers to also  opt-in to preferred data updates not available in their current data subscription tier.

\section{Use Case: Long Distance Hikers}
Our subscription model allows a range of use. We provide an example of how hikers (both as buyers and sellers) would benefit from our service. Long-distance hikers (thru-hikers who hike a trail end-to-end, and section hikers who spend up to months hiking a section of a long trail) experience unique events while ``away'' from typical day-to-day life. For instance, hiking a long trail such as the Appalachian Trail is an arduous task that can take a minimum of three months to complete. Appalachian trail hikers do it for various reasons: for a sense of adventure, to conquer self \cite{berg2015conquer} and to collect data \cite{Bryson99}. The motivation to hike a trail often has a flow-on effect on how a hiker experiences the trail \cite{strayed2012wild}. 

A second consideration around hikers is the case of representation. For instance, there is a need to understand how people with disabilities experience the trail \cite{BlindHike}. The under-representation of these populations creates a demand that could be met if there is an incentive model to encourage willing hikers to undertake the hike to meet the need for data to understand these experiences.

A third consideration is doing work. Technology is not prevalent in the wilderness and is an opportunity for hikers to gather data or test some device, individually or in groups. In this case, data collected or generated by hikers could be used to understand remote terrain, how hikers perform health wise in various locations, or testing equipment and gear. The context of the both the hiker and the environment along with the experience makes for valuable insight that could be packaged and sold to the highest bidder.


\subsection{Preparation (Pre-Hike State of the Field)}
Preparing for a thru-hike is a long-term commitment: future hikers have to make arrangements to take vacation from work, arrange affairs of home while on the trail, purchase trail gear which tend to be expensive, and arrange logistics that are made more complicated with the fact that the future thru-hikers are preparing for the unknown. Regardless of motivation for undertaking the hike, most concerns are valid across hikers of different types \cite{berg2015conquer, Berger2001TripleCrown, Bryson99}. 

Part of planning for eventualities include discussing trail conditions and seeking advice from experienced and/or previous hikers. This is made easier especially on established trails like the Appalachian Trail. Previous, current and future thru-hikers have a central place such as online communities \footnote{\url{https://www.reddit.com/r/AppalachianTrail/}} devoted to the trail, where they can discuss the state of the trail, pose and answer (hypothetical or not) questions about their trail experience. 

Considering different hikers, different data is sought: Miles per day, health maintenance (for people with chronic illnesses and those who take daily medication for instance), safety (and how to maintain it), calories taken, weight lost/gained through the hike, longevity of gear, trail conditions, weather conditions, water sources, re-supply points etc. These data metrics are important to hikers for their personal reasons and are rich with context and meaning to provide a full picture of what the hiker has gone through. This very context is the data gold mine for select buyers especially if you are a well known seller.

There is a set of hikers (mostly those with name recognition brought about by social media) that undertake sponsored hikes. The sponsorships range from an exchange of free gear for an honest review,  to a fully sponsored hike to collect data and provide updates via podcasts. Agreements are typically set at the beginning of the hike, and deliverables planned. Our service democratizes the sponsorship process, allowing for a larger population to be able to participate as sponsors to hikers and/or trails that they are interested in.

\begin{marginfigure}
\includegraphics[width=1\linewidth]{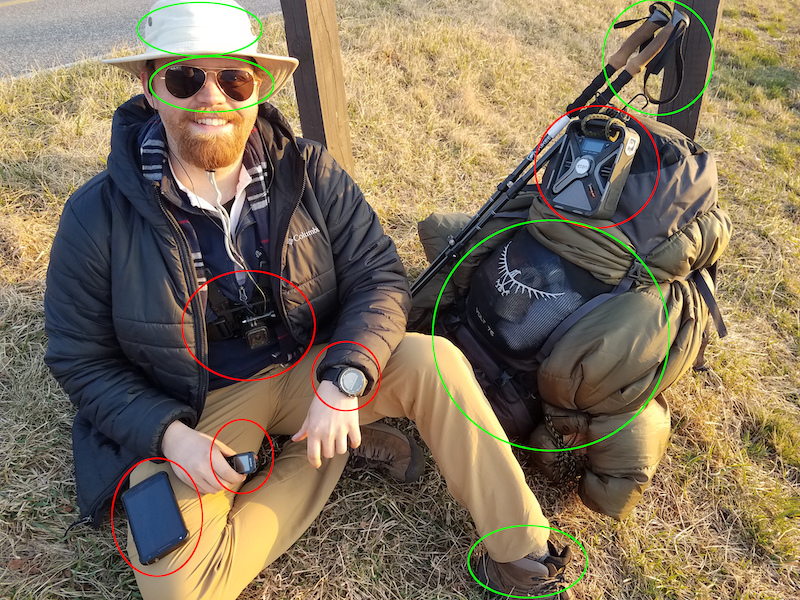}
\centering 
\caption{Figure shows some placement of sensors on the hiker's body which borrows from how hikers currently place and use technology \cite{Stelter17}. The red circles show current technologies placed on the hiker, and the green circles show some opportunities to have additional sensors. Including sponsored sensors.}
\label{fig:sensor-placement}
\end{marginfigure}

\subsection{On-Hike (How Hikers Currently Update)}
Long distance hikers typically update their social media letting their followers know of their progress, to solicit help (typically morale-related, or to ask for rides to resupply towns). The communication medium is chosen by the hiker to fit their persona, ease of use, and privacy. Considering those with updates set to public, those on Instagram typically post multiple times a week, while those on YouTube do so once are week or longer. Frequency of posting typically relies on the availability of connectivity and the pre-processing time, so pictures with brief descriptions are faster to output than videos. Some hikers do both. Our service eases the process of updating by providing connectivity even while on the trail (Figure \ref{fig:sensor-placement}). Too, our dashboard provides a central place where updates are automatically incorporated from various social media and subsequently alerting the buyer about new information depending on the notification triggers (Figure \ref{fig:push-notification}) they set. 

\subsection{Post-Hike (``Patience, Young Grasshopper'')}
Post-hike, long distance hikers become senior members of the hiking community with institutional knowledge about the trail. Their role in the community evolves to pass on their knowledge and wisdom about other hikers they met on the trail, together with their trail experience to others in the community and also potential hikers. The experience gained and elevation gives the hiker a chance  to partake in bigger selling opportunity where another hike could be performed or taking a group of hikers onto the trail (allowing more data to be collected in a co-located context). In essence, the cycle is completed. This is typically where the hikers would share other data and information they experienced on the trail that was either presented from data posts or not (either due to privacy concerns, the processing involved to showcase the data, or lack of connectivity). The difference in the data provided is affected by time and context: while the accuracy of recall is degraded by the passage of time, the richness of information is increased as a factor of nuance and context. 

Our service allows for ease-of-use whenever the hiker is ready to share the data. While our subscription just-in-time service ensures data integrity while the hiker is on the trail, we are nimble enough to allow for data sharing even after the hike is completed. A buyer can evolve to become a seller, or can be a buyer to a different subscription, while being a seller of their own data, allowing for control over personal information. 

As we have shown, data control is the future. We offer incentives to our early adopter of our service. For potential sellers, for a brief-period, we offer \textit{milestone bounties} for your activity. For buyers, we offer 2\% off of the fees from each tier as our thanks for being an early adopter. 

Sign up today!

\section{Author's Statement}

This design fiction probes the notion of what happens when personal data is formally turned into a commodity, instead of the indirect way it is currently done, where the users trade their activity data for access to ``free services'' such as the use of social media accounts. What happens when users control the data collected and what is shared in return for cash? Will over-sharing in exchange for hard currency become the norm, or will it lead users to choose to keep their information private? We use the ``Willing Buyer, Willing Seller'' principle to provide transparency for all parties involved by using the activity of hiking as an example. Our use of design fiction also allows us to critically explore the ethics \cite{Lindley2016} of the future of personal data. 

\subsection{Personal Control}
Part of the frustration with the current state in the field in personal information is the lack of control and the lack of knowledge of what exact information is collected, sold \cite{Simeonovski2017} or used to re-target the user \cite{Dolin2018}. The terms of use that users agree with as a condition for using an application are known to use difficult language and are geared towards protecting the providers rather than the user. They also tend to be modified with/without additional consent requested from the user. We consider the move towards users taking control over their data to be inevitable and ponder what it would mean to have the control over what is shared and what is sold.

\subsection{Extensibility}
The use of the hiker term \textit{``The Trail Provides''} allows us to consider a case where money can used as an incentive to make decisions, such as to undertake (un)planned hikes depending on the amount of money available for incentives. As noted earlier in the paper, we considered the case of incentivizing atypical hikers \cite{Coburn2006} to undertake hikes, or experienced/frequent hikers to hike a specific or unmarked trail. This model can be extended as a dare, allowing users to share information of dare-devils act that is transmitted in real-time opens up a Pandora's box of how the dare-devil culture can be used to incentivize dangerous or illegal tasks.
 
\subsection{Gains and Losses: Data for Sale vs Privacy and Ethics}
With the ``Willing buyer, willing seller'' approach, the buyer is provided a just, market-value compensation for their data. Additionally, they will receive compensation for the life of the data which can also be extended to provide earnings to the heir(s) -- as is currently the case with literature and music publishing. This allows the user to obtain benefits that is otherwise kept from them, given that a lot of data collected from the users by current services are never destroyed, and can be re-sold as long as demand persists. We tackle the issue of data manageability of unwieldy data when discussing our dashboard design, which not only allows for a one-stop collection of data, but also provide  visualization choices, and therefore context of how the data collected is interpreted. 

The use of subscription data also raises an ethical question about ownership. Even though the seller is the owner of data and controls what is sold, and for how much; for having paid for the use of the data, do the buyers have a license to the data? What if the seller exercises their right to be forgotten \cite{rosen2011right} by choosing to no longer make their data available online, what happens to data justly paid for?

\subsection{Discussion}
The \textit{``Willing buyer, willing seller''} principle allows us to first consider the future of data control, and second to consider the ethics of data as a transaction item through the lens of the hiking context. It allows for the discussion on the price of privacy, whether this can be quantified, and the ethics of ownership (For example, do parents own (and therefore receive revenue from) their children's data? Even if the children attain majority?). It also borrows the central thesis from the current state-of-the-art in how companies obtain compensation for selling user data. It is our hope that our use of this principle will provide for an interesting perspective to discuss the ethics of data ownership and the future of data privacy.


\bibliographystyle{ACM-Reference-Format}
\bibliography{fiction-ref} 

\end{document}